\documentclass[]{pasj01}
\Received{}
\Accepted{}
 
\begin{document} 

\title{
Optical follow-up observation for GW event S190510g using Subaru/Hyper Suprime-Cam}


\author{Takayuki \textsc{Ohgami}\altaffilmark{1}}
\email{t-ohgami@konan-u.ac.jp}

\author{Nozomu \textsc{Tominaga}\altaffilmark{1, 2}}
\author{Yousuke \textsc{Utsumi}\altaffilmark{3}}
\author{Yuu \textsc{Niino}\altaffilmark{4, 5}}
\author{Masaomi \textsc{Tanaka}\altaffilmark{6}}
\author{Smaranika \textsc{Banerjee}\altaffilmark{6}}

\author{Ryo \textsc{Hamasaki}\altaffilmark{1}}

\author{Michitoshi \textsc{Yoshida}\altaffilmark{7}}
\author{Tsuyoshi \textsc{Terai}\altaffilmark{7}}
\author{Yuhei \textsc{Takagi}\altaffilmark{7}}

\author{Tomoki \textsc{Morokuma}\altaffilmark{4}}
\author{Mahito \textsc{Sasada}\altaffilmark{8, 9}}
\author{Hiroshi \textsc{Akitaya}\altaffilmark{8}}
\author{Naoki \textsc{Yasuda}\altaffilmark{2}}
\author{Kenshi \textsc{Yanagisawa}\altaffilmark{10, 8, 11}}
\author{Ryou \textsc{Ohsawa}\altaffilmark{4}}

\author{the J-GEM collaboration}

\altaffiltext{1}{Department of Physics, Faculty of Science and Engineering, Konan University, 8-9-1 Okamoto, Kobe, Hyogo 658-8501, Japan}
\altaffiltext{2}{Kavli Institute for the Physics and Mathematics of the Universe (WPI), The University of Tokyo, 5-1-5 Kashiwanoha, Kashiwa, Chiba 277-8583, Japan}
\altaffiltext{3}{Kavli Institute for Particle Astrophysics and Cosmology (KIPAC), SLAC National Accelerator Laboratory, Stanford University, 2575 Sand Hill Road, Menlo Park, CA  94025, USA}
\altaffiltext{4}{Institute of Astronomy, Graduate School of Science, The University of Tokyo, 2-21-1 Osawa, Mitaka, Tokyo 181-0015, Japan}
\altaffiltext{5}{Research Center for the Early Universe, Graduate School of Science, The University of Tokyo, 7-3-1 Hongo, Bunkyo-ku, Tokyo 113-0033, Japan}

\altaffiltext{6}{Astronomical Institute, Tohoku University, Sendai 980-8578, Japan}
\altaffiltext{7}{Subaru Telescope, National Astronomical Observatory of Japan, 650 North A`ohoku Place, Hilo, HI 96720, USA}
\altaffiltext{8}{Hiroshima Astrophysical Science Center, Hiroshima University, 1-3-1 Kagamiyama, Higashi-Hiroshima, Hiroshima 739-8526, Japan}
\altaffiltext{9}{Mizusawa VLBI Observatory, National Astronomical Observatory of Japan, 2-12 Hoshigaoka, Mizusawa, Oshu, Iwate 023-0861, Japan}
\altaffiltext{10}{National Astronomical Observatory of Japan, 2-21-1 Osawa, Mitaka, Tokyo 181-8588, Japan}
\altaffiltext{11}{Department of Astronomy, School of Science, The University of Tokyo, 7-3-1 Hongo, Bunkyo-ku, Tokyo 113-0033, Japan}



\maketitle
\begin{abstract}
A gravitational wave event, S190510g, which was classified as a binary-neutron-star coalescence at the time of preliminary alert, was detected by LIGO/Virgo collaboration on May 10, 2019.
At 1.7 hours after the issue of its preliminary alert, we started a target-of-opportunity imaging observation in $Y$-band to search for its optical counterpart using the Hyper Suprime-Cam (HSC) on the Subaru Telescope.
The observation covers a 118.8 deg$^{2}$ sky area corresponding to 11.6\% confidence in the localization skymap released in the preliminary alert and 1.2\% in the updated skymap.
We divided the observed area into two fields based on the availability of HSC reference images.
For the fields with the HSC reference images, we applied an image subtraction technique; for the fields without the HSC reference images, we sought individual HSC images by matching a catalog of observed objects with the PS1 catalog.
The search depth is 22.28 mag in the former method and the limit of search depth is 21.3 mag in the latter method.
Subsequently, we performed visual inspection and obtained 83 candidates using the former method and 50 candidates using the latter method.
Since we have only the 1-day photometric data, we evaluated probability to be located inside the 3D skymap by estimating their distances with photometry of associated extended objects. We found three candidates are likely located inside the 3D skymap and concluded they could be an counterpart of S190510g, while most of 133 candidates were likely to be supernovae because the number density of candidates was consistent with the expected number of supernova detections.
By comparing our observational depth with a light curve model of such a kilonova reproducing AT2017gfo, we show that early-deep observations with the Subaru/HSC can capture the rising phase of blue component of kilonova at the estimated distance of S190510g ($\sim$230 Mpc).
\end{abstract}

\section{Introduction}
A multi-messenger observation with gravitational waves (GW) and electromagnetic (EM) waves is crucial for understanding physical processes of compact star coalescence.
Neutron-star (NS) mergers are expected to be accompanied by EM emissions called ``kilonova'' (or ``macronova'') powered by radioactive decays of $r$-process nuclei (\cite{1998ApJ...507L..59L, Kulkarni:2005jw, 2010MNRAS.406.2650M}); therefore, the EM emission from BNS-merger events aids in understanding the origin of heavy elements produced by the $r$-process (\cite{2010MNRAS.406.2650M, 2013ApJ...774...25K, 2013ApJ...775...18B, 2013ApJ...775..113T, 2014ApJ...780...31T, 2015MNRAS.450.1777K}).

The localization area of GW observations can be an order of 10 deg$^2$ for the best case, but can be as large as 1000 deg$^2$.
It has been quite large for locating a galaxy hosting a system that caused the GW event.
Therefore, EM follow-up observations had been expected to play a key role in identifying the counterpart.
The first identification of EM counterpart to GW was achieved in the event of the first detection of GW from a neutron star merger (GW170817).
GW170817 was localized with three interferometers in the second observing run (O2) of the LIGO/Virgo collaboration (\cite{TheLIGOScientific:2017qsa}).
The identification of the EM counterpart was made by several observatories on earth including space from radio to gamma ray (\cite{Arcavi:2017vbi, 2017Sci...358.1556C, Diaz:2017uch, Evans:2017mmy, 2017ApJ...850L...1L, Soares-Santos:2017lru, Tanvir:2017pws, Tominaga:2017cgo, 2017ApJ...848L..24V}).

Untargeted wide-field surveys are important for identifying the uniqueness of the counterpart.
The Japanese collaboration for Gravitational wave ElectroMagnetic follow-up (J-GEM; \cite{2016PASJ...68L...9M}) conducted coordinated observations (\cite{Utsumi:2017cti}) and deep blind $z$-band imaging surveys to identify an EM counterpart using Hyper Suprime-Cam (HSC) on the Subaru Telescope (\cite{2018PASJ...70S...1M, 2018PASJ...70...66K, 2018PASJ...70S...2K, 2018PASJ...70S...3F}). They succeeded in independently identifying the counterpart (AT2017gfo; \cite{Tominaga:2017cgo}).
HSC is a 1.5 deg $\phi$ wide-field optical imager, which is the largest among the current existing telescopes with an aperture larger than 8 m.
While galaxy-targeted and untargeted wide-field surveys identified AT2017gfo, wide-field survey observations with the Subaru/HSC and Blanco/Dark Energy Camera (DECam) succeeded in identifying the uniqueness of AT2017gfo with a high completeness by ruling out the other candidates including transients which are not associated with galaxy.

Kilonova models can broadly reproduce the time evolution of optical and near-infrared emissions of AT2017gfo (\cite{2017PhRvD..96l3012S, Tanaka:2017qxj}; \cite{2017Natur.551...80K, Perego_2017, Kawaguchi_2018, Rosswog_2018}).
However, the observed emissions display blue components in the early-phase spectra, and the origin of the emission is unclear.
Two models for the early blue component are proposed: radioactive heating model (a kilonova model having higher electron fraction \cite{Tanaka:2017qxj, 2017ApJ...851L..21V, 2018MNRAS.481.3423W}) and shock heating model (a cocoon emission model \cite{2017Sci...358.1559K, 2018ApJ...855..103P}).
These models can reproduce the EM emission after 0.5 days from the explosion, at which the first observation of AT2017gfo was performed.
They predict different behaviors that predate 0.5 days from the explosion (\cite{Arcavi_2018}), i.e. a cocoon model shows a higher luminosity than the radioactive kilonova model; therefore, the earlier observations for future events are important to discriminate these models.

The LIGO/Virgo collaboration started their third observation run (O3) in April 2019.
They detected a BNS event named GW190425 (\cite{2020ApJ...892L...3A}) on April 25, 2019 at 08:18:05 UTC (GCN, \cite{GCN:24167}) for the first time in O3.
On May 10, 2019 at 02:59:39 UTC, they detected the third BNS event in O3, S190510g, using three interferometers (GCN, \cite{GCN:24442}).
They analyzed the GW signal using BAYESTAR pipeline (\cite{Singer:2015ema}) and released a preliminary localization skymap on May 10, 2019 at 04:03:45 UTC.
The 50\% and 90\% confidence regions correspond to the areas of 575 deg$^{2}$ and 3462 deg$^{2}$, respectively.
The luminosity distance was $269\pm108$ Mpc.
In this alert, the GW event was classified as a BNS coalescence with 98\% confidence level and a false alarm rate (FAR) of $8.4\times10^{-10}$ Hz (about one in 37 years).

On receiving this alert, we conducted a target of opportunity (ToO) imaging observation (GCN, \cite{GCN:24450}), which covered 118.8 deg$^{2}$ corresponding to the integrated probability of 11.6\% in the localization skymap, using the Subaru/HSC.
After our observations, we received an improved localization skymap which is reanalyzed with the LALInference pipeline (\cite{Veitch:2014wba}) by the LIGO/Virgo collaboration on May 10, 2019 at 10:06:59 UTC (GCN, \cite{GCN:24448}).
The 90\% localization area and the luminosity distance were revised to 1166 deg$^2$ and $227\pm92$ Mpc, respectively.
The integrated probability in our observation area decreased to 1.2\% of the total probability owing to the revision.
In this alert, the probability of the event being a BNS-merger event decreased to 42\% (the probability of it being a terrestrial event increased to 58\%) with an FAR of $8.8\times 10^{-9}$ Hz (about one in 3.6 years).

In this paper, we describe the details of the observation of GW event S190510g using the Subaru/HSC, the candidate selection, and a list of candidates.
We investigate the nature of the candidates by estimating a contamination from supernovae.
Finally, we discuss the future prospects for optical-follow-up observations using Subaru/HSC.
In this paper, all magnitudes are given as AB magnitudes.

\section{Observation and data analysis}
We commenced a follow-up observation for the GW event S190510g using Subaru/HSC on May 10, 2019 at 05:46:27 UTC, 1 h 43 min after the issue of the preliminary-alert and 2 h 47 min after the GW detection.
Our original plan was to perform $i$- and $z$-band observations for the GW follow-up; however, we conducted the observation in $Y$-band for this event because only the $Y$-band filter was available that night.
We selected 120 healpix grids with high probabilities in the BAYESTAR localization skymap with a HEALPix resolution of \texttt{NSIDE} $= 64$, which corresponds to 0.84 deg$^2$/pix$^2$, allowing the field of views (FoVs) to overlap each other.
The HSC pointings were set as the central coordinates of the each grid (Table \ref{tab:pointings}).
We exposed the 120 pointings with 30 s each and revisited them with a 1-arcmin offset in each pointing at least one hour apart.
The exposure time was determined considering the observation time of the half night and an exposure interval of approximately 34 s (\cite{2012SPIE.8446E..62U}).

\begin{figure*}
	\begin{center}
		\includegraphics[width=\hsize]{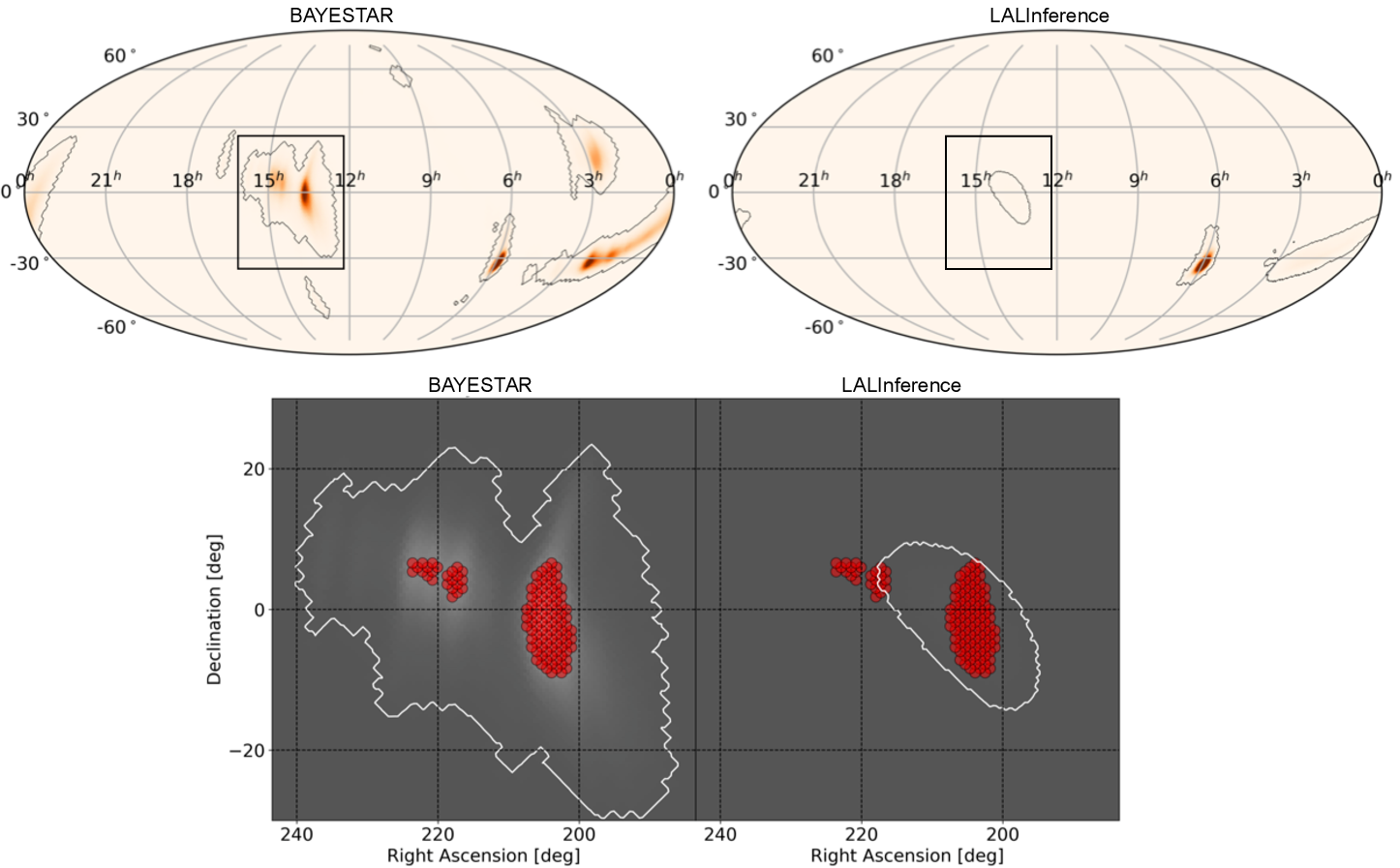} 
	\end{center}
	\caption{Preliminary ({\it top left}, BAYESTAR; GCN, \cite{GCN:24442}) and updated ({\it top right}, LALInference; GCN, \cite{GCN:24448}) localization skymaps of S190510g. White contour lines and red-filled circles ({\it bottom}) represent the localization area corresponding to the 90\% confidence region and our survey pointings, respectively. The observed area of 118.8 deg$^{2}$ corresponds to the 11.6\% and 1.2\% of the total probabilities in the BAYESTAR and LALInference skymaps, respectively.}
	\label{fig:pointings}
\end{figure*}

The survey pointings and 90\% contour for the BAYESTAR skymap are shown in the bottom left panel of Fig. \ref{fig:pointings}.
The observed area of 118.8 deg$^2$ corresponds to the integrated probability of 11.6\% in the BAYESTAR localization skymap.
The skymap is revised significantly to the LALInference localization skymap.
In this updated skymap, the integrated probability in the observed area decreases to 1.2\%.
The 90\% contour in the updated skymap is shown in the right panel of Fig. \ref{fig:pointings}.

\begin{longtable}{ccccc}
	\caption{Central coordinates of the survey pointings and observation log.} \label{tab:pointings}
	\hline
	Pointing & R.A. & decl. & \multicolumn{2}{c}{taiObs} \\
	(ID) & (J2000) & (J2000) & \multicolumn{2}{c}{(UTC)}\\
	\hline \hline
	\endfirsthead
	\hline
	Pointing & R.A. & decl. & \multicolumn{2}{c}{taiObs} \\
	(ID) & (J2000) & (J2000) & \multicolumn{2}{c}{(UTC)}\\
	\hline \hline
	\endhead
	\hline
	\endfoot
	\endlastfoot
	000 & $13^h27^m15^s.06$ & $-08^{\circ}22'52''.4$ & 2019-05-10T05:47:04 & 2019-05-10T07:50:37\\
	001 & $13^h29^m55^s.99$ & $-08^{\circ}59'37''.0$ & 2019-05-10T05:48:05 & 2019-05-10T07:51:38\\
	002 & $13^h27^m15^s.05$ & $-07^{\circ}10'35''.0$ & 2019-05-10T05:49:07 & 2019-05-10T07:52:39\\
	003 & $13^h30^m03^s.80$ & $-07^{\circ}46'42''.3$ & 2019-05-10T05:50:07 & 2019-05-10T07:53:39\\
	004 & $13^h32^m44^s.75$ & $-08^{\circ}23'23''.4$ & 2019-05-10T05:51:08 & 2019-05-10T07:54:40\\
	005 & $13^h35^m41^s.31$ & $-08^{\circ}59'05''.9$ & 2019-05-10T05:52:09 & 2019-05-10T07:55:41\\
	006 & $13^h24^m26^s.29$ & $-05^{\circ}22'30''.1$ & 2019-05-10T05:53:10 & 2019-05-10T07:56:42\\
	007 & $13^h27^m15^s.04$ & $-05^{\circ}58'29''.4$ & 2019-05-10T05:54:10 & 2019-05-10T07:57:43\\
	008 & $13^h29^m56^s.02$ & $-06^{\circ}35'01''.9$ & 2019-05-10T05:55:13 & 2019-05-10T07:58:44\\
	009 & $13^h32^m44^s.77$ & $-07^{\circ}11'06''.1$ & 2019-05-10T05:56:14 & 2019-05-10T07:59:45\\
	010 & $13^h35^m33^s.51$ & $-07^{\circ}47'13''.2$ & 2019-05-10T05:57:15 & 2019-05-10T08:00:46\\
	011 & $13^h38^m30^s.06$ & $-08^{\circ}22'52''.3$ & 2019-05-10T05:58:16 & 2019-05-10T08:01:46\\
	012 & $13^h24^m26^s.28$ & $-04^{\circ}10'37''.9$ & 2019-05-10T05:59:17 & 2019-05-10T08:02:47\\
	013 & $13^h27^m07^s.28$ & $-04^{\circ}47'04''.2$ & 2019-05-10T06:00:20 & 2019-05-10T08:03:48\\
	014 & $13^h30^m03^s.79$ & $-05^{\circ}22'30''.2$ & 2019-05-10T06:01:22 & 2019-05-10T08:04:48\\
	015 & $13^h32^m44^s.77$ & $-05^{\circ}59'00''.5$ & 2019-05-10T06:02:23 & 2019-05-10T08:05:50\\
	016 & $13^h35^m33^s.52$ & $-06^{\circ}35'01''.9$ & 2019-05-10T06:03:23 & 2019-05-10T08:06:51\\
	017 & $13^h38^m30^s.05$ & $-07^{\circ}10'35''.2$ & 2019-05-10T06:04:24 & 2019-05-10T08:07:51\\
	018 & $13^h41^m18^s.80$ & $-07^{\circ}46'42''.2$ & 2019-05-10T06:05:24 & 2019-05-10T08:08:52\\
	019 & $13^h24^m26^s.27$ & $-02^{\circ}58'52''.2$ & 2019-05-10T06:06:30 & 2019-05-10T08:09:54\\
	020 & $13^h27^m07^s.28$ & $-03^{\circ}35'15''.5$ & 2019-05-10T06:07:30 & 2019-05-10T08:10:55\\
	021 & $13^h29^m56^s.03$ & $-04^{\circ}11'09''.0$ & 2019-05-10T06:08:31 & 2019-05-10T08:11:55\\
	022 & $13^h32^m44^s.78$ & $-04^{\circ}47'04''.2$ & 2019-05-10T06:09:32 & 2019-05-10T08:12:56\\
	023 & $13^h35^m41^s.28$ & $-05^{\circ}22'30''.3$ & 2019-05-10T06:10:47 & 2019-05-10T08:13:57\\
	024 & $13^h38^m22^s.27$ & $-05^{\circ}59'00''.5$ & 2019-05-10T06:11:48 & 2019-05-10T08:14:58\\
	025 & $13^h41^m18^s.79$ & $-06^{\circ}34'30''.9$ & 2019-05-10T06:12:49 & 2019-05-10T08:15:58\\
	026 & $13^h44^m07^s.55$ & $-07^{\circ}10'35''.1$ & 2019-05-10T06:13:49 & 2019-05-10T08:17:00\\
	027 & $13^h27^m15^s.02$ & $-02^{\circ}23'01''.3$ & 2019-05-10T06:14:50 & 2019-05-10T08:18:03\\
	028 & $13^h29^m56^s.04$ & $-02^{\circ}59'23''.3$ & 2019-05-10T06:15:51 & 2019-05-10T08:19:04\\
	029 & $13^h32^m52^s.52$ & $-03^{\circ}34'44''.4$ & 2019-05-10T06:16:53 & 2019-05-10T08:20:05\\
	030 & $13^h35^m41^s.27$ & $-04^{\circ}10'38''.0$ & 2019-05-10T06:17:54 & 2019-05-10T08:21:06\\
	031 & $13^h38^m22^s.28$ & $-04^{\circ}47'04''.2$ & 2019-05-10T06:18:54 & 2019-05-10T08:22:07\\
	032 & $13^h41^m11^s.03$ & $-05^{\circ}23'01''.2$ & 2019-05-10T06:19:55 & 2019-05-10T08:23:07\\
	033 & $13^h43^m59^s.77$ & $-05^{\circ}59'00''.4$ & 2019-05-10T06:20:56 & 2019-05-10T08:24:08\\
	034 & $13^h27^m07^s.28$ & $-01^{\circ}11'53''.0$ & 2019-05-10T06:21:57 & 2019-05-10T08:25:13\\
	035 & $13^h30^m03^s.77$ & $-01^{\circ}47'11''.1$ & 2019-05-10T06:22:57 & 2019-05-10T08:26:14\\
	036 & $13^h32^m44^s.78$ & $-02^{\circ}23'32''.4$ & 2019-05-10T06:23:58 & 2019-05-10T08:27:15\\
	037 & $13^h35^m33^s.54$ & $-02^{\circ}59'23''.3$ & 2019-05-10T06:24:59 & 2019-05-10T08:28:15\\
	038 & $13^h38^m22^s.28$ & $-03^{\circ}35'15''.4$ & 2019-05-10T06:26:00 & 2019-05-10T08:29:16\\
	039 & $13^h41^m11^s.03$ & $-04^{\circ}11'09''.0$ & 2019-05-10T06:27:01 & 2019-05-10T08:30:17\\
	040 & $13^h44^m07^s.53$ & $-04^{\circ}46'33''.2$ & 2019-05-10T06:28:02 & 2019-05-10T08:31:18\\
	041 & $13^h27^m07^s.29$ & $-00^{\circ}00'15''.5$ & 2019-05-10T06:29:03 & 2019-05-10T08:32:26\\
	042 & $13^h46^m56^s.28$ & $-05^{\circ}22'30''.1$ & 2019-05-10T06:30:05 & 2019-05-10T08:33:36\\
	043 & $13^h30^m03^s.77$ & $-00^{\circ}35'33''.0$ & 2019-05-10T06:31:06 & 2019-05-10T08:34:45\\
	044 & $13^h32^m52^s.52$ & $-01^{\circ}11'21''.9$ & 2019-05-10T06:32:07 & 2019-05-10T08:35:46\\
	045 & $13^h35^m41^s.27$ & $-01^{\circ}47'11''.3$ & 2019-05-10T06:33:07 & 2019-05-10T08:36:47\\
	046 & $13^h38^m30^s.02$ & $-02^{\circ}23'01''.3$ & 2019-05-10T06:34:08 & 2019-05-10T08:37:48\\
	047 & $13^h41^m11^s.04$ & $-02^{\circ}59'23''.3$ & 2019-05-10T06:35:09 & 2019-05-10T08:38:49\\
	048 & $13^h43^m59^s.78$ & $-03^{\circ}35'15''.4$ & 2019-05-10T06:36:09 & 2019-05-10T08:39:49\\
	049 & $13^h46^m48^s.53$ & $-04^{\circ}11'09''.0$ & 2019-05-10T06:37:10 & 2019-05-10T08:40:50\\
	050 & $13^h30^m03^s.76$ & $-00^{\circ}35'33''.1$ & 2019-05-10T06:38:11 & 2019-05-10T08:42:00\\
	051 & $13^h32^m44^s.79$ & $-00^{\circ}00'15''.4$ & 2019-05-10T06:39:13 & 2019-05-10T08:43:02\\
	052 & $13^h35^m41^s.26$ & $-00^{\circ}35'33''.1$ & 2019-05-10T06:40:14 & 2019-05-10T08:44:03\\
	053 & $13^h38^m22^s.29$ & $-01^{\circ}11'53''.0$ & 2019-05-10T06:41:15 & 2019-05-10T08:45:03\\
	054 & $13^h41^m11^s.04$ & $-01^{\circ}47'42''.3$ & 2019-05-10T06:42:15 & 2019-05-10T08:46:04\\
	055 & $13^h29^m56^s.03$ & $+01^{\circ}47'11''.3$ & 2019-05-10T06:43:16 & 2019-05-10T08:47:11\\
	056 & $13^h44^m07^s.52$ & $-02^{\circ}23'01''.4$ & 2019-05-10T06:44:17 & 2019-05-10T08:48:21\\
	057 & $13^h46^m48^s.53$ & $-02^{\circ}59'23''.3$ & 2019-05-10T06:45:18 & 2019-05-10T08:49:22\\
	058 & $13^h32^m52^s.52$ & $+01^{\circ}11'53''.0$ & 2019-05-10T06:46:18 & 2019-05-10T08:50:32\\
	059 & $13^h35^m41^s.27$ & $-00^{\circ}35'33''.0$ & 2019-05-10T06:47:19 & 2019-05-10T08:51:33\\
	060 & $13^h38^m30^s.02$ & $+00^{\circ}00'15''.6$ & 2019-05-10T06:48:20 & 2019-05-10T08:52:34\\
	061 & $13^h30^m03^s.77$ & $+02^{\circ}59'23''.3$ & 2019-05-10T06:49:21 & 2019-05-10T08:53:40\\
	062 & $13^h41^m11^s.04$ & $-00^{\circ}36'04''.0$ & 2019-05-10T06:50:22 & 2019-05-10T08:54:49\\
	063 & $13^h43^m59^s.79$ & $-01^{\circ}11'53''.0$ & 2019-05-10T06:51:23 & 2019-05-10T08:55:50\\
	064 & $13^h32^m52^s.52$ & $+02^{\circ}23'32''.4$ & 2019-05-10T06:52:24 & 2019-05-10T08:57:00\\
	065 & $13^h46^m56^s.27$ & $-01^{\circ}47'11''.2$ & 2019-05-10T06:53:25 & 2019-05-10T08:58:12\\
	066 & $13^h49^m45^s.02$ & $-02^{\circ}23'01''.2$ & 2019-05-10T06:54:26 & 2019-05-10T08:59:13\\
	067 & $13^h35^m33^s.53$ & $+01^{\circ}47'11''.3$ & 2019-05-10T06:55:27 & 2019-05-10T09:00:25\\
	068 & $13^h38^m30^s.02$ & $+01^{\circ}11'53''.0$ & 2019-05-10T06:56:28 & 2019-05-10T09:01:26\\
	069 & $13^h41^m18^s.76$ & $-00^{\circ}35'33''.1$ & 2019-05-10T06:57:29 & 2019-05-10T09:02:26\\
	070 & $13^h32^m52^s.53$ & $+03^{\circ}35'15''.7$ & 2019-05-10T06:58:29 & 2019-05-10T09:03:35\\
	071 & $13^h44^m07^s.52$ & $+00^{\circ}00'15''.7$ & 2019-05-10T06:59:30 & 2019-05-10T09:04:46\\
	072 & $13^h46^m56^s.27$ & $-00^{\circ}35'33''.0$ & 2019-05-10T07:00:31 & 2019-05-10T09:05:47\\
	073 & $13^h35^m33^s.53$ & $+02^{\circ}58'52''.4$ & 2019-05-10T07:01:32 & 2019-05-10T09:06:58\\
	074 & $13^h49^m37^s.29$ & $-01^{\circ}11'53''.0$ & 2019-05-10T07:02:33 & 2019-05-10T09:08:12\\
	075 & $13^h38^m22^s.28$ & $+02^{\circ}23'01''.4$ & 2019-05-10T07:03:33 & 2019-05-10T09:09:22\\
	076 & $13^h32^m52^s.53$ & $+04^{\circ}47'04''.2$ & 2019-05-10T07:04:34 & 2019-05-10T09:10:28\\
	077 & $13^h41^m11^s.04$ & $+01^{\circ}47'11''.3$ & 2019-05-10T07:05:35 & 2019-05-10T09:11:37\\
	078 & $13^h43^m59^s.79$ & $+01^{\circ}11'22''.0$ & 2019-05-10T07:06:36 & 2019-05-10T09:12:38\\
	079 & $13^h35^m41^s.28$ & $+04^{\circ}11'09''.1$ & 2019-05-10T07:07:36 & 2019-05-10T09:13:47\\
	080 & $13^h46^m56^s.26$ & $-00^{\circ}35'33''.1$ & 2019-05-10T07:08:37 & 2019-05-10T09:15:01\\
	081 & $13^h49^m37^s.29$ & $-00^{\circ}00'15''.5$ & 2019-05-10T07:09:38 & 2019-05-10T09:16:02\\
	082 & $13^h38^m30^s.02$ & $+03^{\circ}35'15''.5$ & 2019-05-10T07:10:39 & 2019-05-10T09:17:14\\
	083 & $13^h32^m44^s.77$ & $+05^{\circ}58'29''.5$ & 2019-05-10T07:11:40 & 2019-05-10T09:18:21\\
	084 & $13^h41^m11^s.04$ & $+02^{\circ}58'52''.3$ & 2019-05-10T07:12:41 & 2019-05-10T09:19:32\\
	085 & $13^h35^m33^s.52$ & $+05^{\circ}22'30''.3$ & 2019-05-10T07:13:42 & 2019-05-10T09:20:38\\
	086 & $13^h43^m59^s.79$ & $+02^{\circ}23'01''.4$ & 2019-05-10T07:14:43 & 2019-05-10T09:21:49\\
	087 & $13^h46^m48^s.54$ & $+01^{\circ}47'11''.3$ & 2019-05-10T07:15:44 & 2019-05-10T09:22:50\\
	088 & $13^h38^m22^s.27$ & $+04^{\circ}46'33''.2$ & 2019-05-10T07:16:45 & 2019-05-10T09:24:00\\
	089 & $13^h41^m18^s.77$ & $+04^{\circ}11'09''.1$ & 2019-05-10T07:17:45 & 2019-05-10T09:25:01\\
	090 & $13^h35^m41^s.30$ & $+06^{\circ}35'02''.2$ & 2019-05-10T07:18:47 & 2019-05-10T09:26:08\\
	091 & $13^h44^m07^s.52$ & $+03^{\circ}35'15''.6$ & 2019-05-10T07:19:48 & 2019-05-10T09:27:20\\
	092 & $13^h38^m22^s.26$ & $+05^{\circ}58'29''.5$ & 2019-05-10T07:20:48 & 2019-05-10T09:28:26\\
	093 & $13^h46^m48^s.54$ & $+02^{\circ}58'52''.3$ & 2019-05-10T07:21:49 & 2019-05-10T09:29:38\\
	094 & $13^h41^m18^s.79$ & $+05^{\circ}23'01''.5$ & 2019-05-10T07:22:49 & 2019-05-10T09:30:44\\
	095 & $13^h44^m07^s.53$ & $+04^{\circ}47'04''.3$ & 2019-05-10T07:23:50 & 2019-05-10T09:31:46\\
	096 & $14^h31^m56^s.27$ & $+01^{\circ}47'42''.4$ & 2019-05-10T07:24:57 & 2019-05-10T09:33:49\\
	097 & $14^h29^m07^s.51$ & $+02^{\circ}23'32''.4$ & 2019-05-10T07:25:58 & 2019-05-10T09:34:50\\
	098 & $14^h26^m18^s.77$ & $+02^{\circ}59'23''.4$ & 2019-05-10T07:26:59 & 2019-05-10T09:35:51\\
	099 & $14^h31^m48^s.53$ & $+02^{\circ}58'52''.3$ & 2019-05-10T07:28:00 & 2019-05-10T09:36:52\\
	100 & $14^h28^m59^s.78$ & $+03^{\circ}34'44''.5$ & 2019-05-10T07:29:00 & 2019-05-10T09:37:52\\
	101 & $14^h26^m11^s.03$ & $+04^{\circ}10'38''.0$ & 2019-05-10T07:30:01 & 2019-05-10T09:38:53\\
	102 & $14^h34^m45^s.02$ & $+03^{\circ}35'15''.6$ & 2019-05-10T07:31:03 & 2019-05-10T09:39:58\\
	103 & $14^h31^m56^s.27$ & $+04^{\circ}11'09''.1$ & 2019-05-10T07:32:03 & 2019-05-10T09:40:59\\
	104 & $14^h26^m18^s.78$ & $+05^{\circ}23'01''.4$ & 2019-05-10T07:33:04 & 2019-05-10T09:42:00\\
	105 & $14^h29^m07^s.53$ & $+04^{\circ}47'04''.3$ & 2019-05-10T07:34:05 & 2019-05-10T09:43:01\\
	106 & $14^h34^m37^s.28$ & $+04^{\circ}46'33''.2$ & 2019-05-10T07:35:06 & 2019-05-10T09:44:02\\
	107 & $14^h28^m59^s.77$ & $+05^{\circ}58'29''.5$ & 2019-05-10T07:36:07 & 2019-05-10T09:45:04\\
	108 & $14^h31^m56^s.28$ & $+05^{\circ}23'01''.4$ & 2019-05-10T07:37:08 & 2019-05-10T09:46:05\\
	109 & $14^h43^m11^s.28$ & $+04^{\circ}11'09''.0$ & 2019-05-10T07:38:08 & 2019-05-10T09:47:15\\
	110 & $14^h46^m00^s.03$ & $+04^{\circ}47'04''.2$ & 2019-05-10T07:39:09 & 2019-05-10T09:48:16\\
	111 & $14^h43^m03^s.52$ & $+05^{\circ}22'30''.3$ & 2019-05-10T07:40:10 & 2019-05-10T09:49:17\\
	112 & $14^h40^m14^s.77$ & $+05^{\circ}58'29''.6$ & 2019-05-10T07:41:10 & 2019-05-10T09:50:18\\
	113 & $14^h48^m41^s.02$ & $+05^{\circ}22'30''.4$ & 2019-05-10T07:42:11 & 2019-05-10T09:51:24\\
	114 & $14^h46^m00^s.04$ & $+05^{\circ}59'00''.6$ & 2019-05-10T07:43:12 & 2019-05-10T09:52:25\\
	115 & $14^h43^m11^s.29$ & $+06^{\circ}35'02''.2$ & 2019-05-10T07:44:13 & 2019-05-10T09:53:26\\
	116 & $14^h54^m18^s.52$ & $+05^{\circ}22'30''.3$ & 2019-05-10T07:45:14 & 2019-05-10T09:54:37\\
	117 & $14^h51^m29^s.77$ & $+05^{\circ}58'29''.5$ & 2019-05-10T07:46:15 & 2019-05-10T09:55:38\\
	118 & $14^h48^m48^s.79$ & $+06^{\circ}35'02''.1$ & 2019-05-10T07:47:15 & 2019-05-10T09:56:39\\
	119 & $14^h54^m18^s.52$ & $+06^{\circ}34'31''.0$ & 2019-05-10T07:48:16 & 2019-05-10T09:57:41\\
	\hline
\end{longtable}

We reduce the observational data with {\it hscPipe} v4.0.5 (\cite{2018PASJ...70S...5B}).
This pipeline is a standard reduction pipeline for HSC and provides complete packages for the analyses of image data, including bias subtraction, flat fielding, astrometry, flux calibration, mosaicing, warping, stacking, image subtraction, source detection, and source measurement.
We estimate 5$\sigma$ limiting magnitudes in the single-exposure images by measuring standard deviations of sky fluxes in randomly distributed apertures with a diameter of twice the full width at half maximum (FWHM) of a point spread function (PSF).
Subsequently, we found that the $Y$-band limiting magnitudes have a mode value of 22.30 mag.
The mode value of the seeings in each single-exposure image is approximately 0.6 arcsec.

\section{Selection methods for transient objects and results}
\begin{figure}
	\begin{center}
		\includegraphics[width=5.cm]{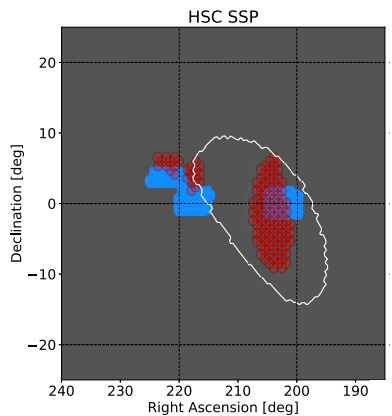} 
	\end{center}
	\caption{Coverage of the deep Subaru/HSC reference images. Blue area shows the footprint of the reference images we used. Red-filled circles represent our observation pointings.}
	\label{fig:RefCoverage}
\end{figure}

In this section, we describe the methods used to search for transient objects related to S190510g and the corresponding results.
We apply the following two methods:
One is to search for transients with an image-subtraction technique, which is useful for searching transient objects embedded in host galaxies.
However, deep reference images are not available for all the survey pointings as shown in Fig. \ref{fig:RefCoverage}.
The area with deep Subaru/HSC reference images is 25.9 deg$^2$ while the area without deep reference image is 92.9 deg$^2$.
Therefore, we adopt the second method for the remaining area.
This method searches the single-exposure images without using the image subtraction by matching known objects in the Pan-STARRS1 (PS1; \cite{2016arXiv161205243F}) catalog.

\subsection{Selection in fields with HSC-SSP reference images}
First, we apply image subtraction for the fields with deep reference images.
We use images obtained in the HSC Subaru Strategic Program (SSP; \cite{2018PASJ...70S...8A}) as the reference images.
These reference images were taken from March 25, 2014 to April 8, 2019, and the total exposure time for each field is 200 s.
The limiting magnitude of reference images is 23.3 mag according to the HSC Exposure Time Calculator\footnote{https://hscq.naoj.hawaii.edu/cgi-bin/HSC\_ETC/hsc\_etc.cgi} (ETC), and is substantially deeper than our observations.
Thus, the search depth in the subtracted images are determined by our images because they dominate the noise of the subtracted images.
The subtraction package in the {\it hscPipe} implementation is based on an algorithm proposed in \citet{1998ApJ...503..325A} and \citet{1999astro.ph..3111A}.
The seeings in the reference images are blurred to those in our observation images by being convolved with kernels to make their PSFs equivalent in this algorithm.
The mode value of 5$\sigma$ limiting magnitudes for the images subtracted with the reference images is 22.28 mag, obtained using the same method as described in Section 2.
The depth is not degraded even after the subtraction because the reference images are sufficiently deep.

\begin{figure}
	\begin{center}
		\includegraphics[width=8.cm]{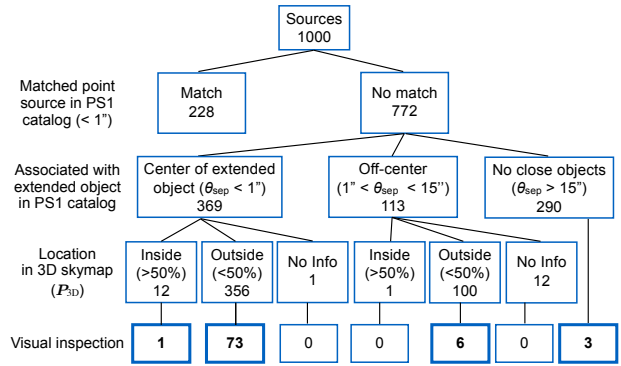}
	\end{center}
	\caption{Flowchart of the candidate screening and the classification process for the objects in the difference images. Numbers in each box represent the number of remaining objects at each step. Thick boxes indicate the candidates after the visual inspection.}
	\label{fig:flowchart_SSP}
\end{figure}

We then perform a candidate selection in the different images to exclude bogus detections (e.g., caused by bad pixels in the reference images or failure of the image subtraction) and moving objects by referring to the criteria used in \citet{Tominaga:2017cgo}.
First, we set the following five criteria; (i) $|(S/N)_{\rm PSF}|>5$, (ii) $(b/a)/(b/a)_{\rm PSF}>0.65$ where $a$ and $b$ are the lengths of major and minor axes of the shape of an object, respectively (iii) $0.7<{\rm FWHM}/({\rm FWHM})_{\rm PSF}<1.3$, (iv) PSF-subtracted residual $< 3\sigma$-standard-deviation range (in the difference image), and (v) detected at least twice and one hour apart.
The criteria (ii) and (iii) are adopted to identify point sources, and (iv) is required to confirm that the objects can be described by a PSF.
The criterion (v) is applied to exclude moving objects, such as minor planets.
We find 1000 objects satisfying these criteria.
Figure \ref{fig:flowchart_SSP} shows a flowchart of the selection process for these objects.
Since our observation area corresponds to the footprints of the PS1 survey (\cite{2016arXiv161205560C}), we match these objects with the PS1 catalog (\cite{2016arXiv161205243F}) to exclude objects associated with known stellar-like objects (point source) within 1 arcsec.
To classify the point sources, we use a flag of extended in \texttt{objInfoFlag} in the PS1 catalog.
By this process, 228 objects are excluded.

Next, we classify the remaining 772 objects by their angular separation $\theta_{\rm sep}$ from the nearby extended objects.
We use the catalog of extended objects taken from the PS1 catalog.
We obtain 369 objects located at the center of an extended object ($\theta_{\rm sep}<1''$), 113 objects located at off-center ($1''<\theta_{\rm sep}<15''$), and 290 objects that have no close extended objects ($\theta_{\rm sep}>15''$).
Moreover, we statistically evaluate the probability of the associated extended objects inside the 3D localization map using the observed magnitudes and the luminosity function of galaxies.
We calculate a probability $P_{\rm 3D}$ that an extended object is located inside a 3$\sigma$ range of the LALInference 3D skymap following the method described in \citet{Tominaga:2017cgo}.
The $P_{\rm 3D}$ is defined as follows:
\begin{eqnarray}
	P_{\rm 3D}(\lambda_{j},\,m_{j}):=\frac{\int^{D_{\rm mean}+3\sigma_{D}}_{D_{\rm mean}-3\sigma_{D}} \phi\, A \,dD}{\int^{\infty}_{0}\phi\, A\,dD},
	\label{eq:P3D}
\end{eqnarray}
where $\phi = \phi(\lambda,\,M)$ is the luminosity function of galaxies at a rest wavelength $\lambda$ derived from the rest-frame $UBVRI$-luminosity functions (\cite{2005A&A...439..863I}) and the {\it Planck} cosmology (\cite{2014A&A...571A..16P}), $A = A(D)$ is an observed surface area at a distance of $D$, $D_{\rm mean}$ and $\sigma_{D}$ are mean value and standard deviation of a probability distribution of the distance, respectively, $M = M(D;\,m_{j})$ is an absolute magnitude of a galaxy with observer-frame $j$-band apparent magnitude $m_{j}$ at a distance of $D$, and $\lambda = \lambda(D;\,\lambda_{j})$ is the rest wavelength redshifted from a observed wavelength $\lambda_{j}$ at a distance of $D$.
We correct the Galactic extinction (\cite{2011ApJ...737..103S})\footnote{http://irsa.ipac.caltech.edu/applications/DUST/} when we convert the apparent magnitudes to the absolute magnitudes.
We use $r$- and/or $i$-band PSF-magnitudes (\texttt{rMeanPSFMag}, \texttt{iMeanPSFMag} in PS1 catalog) of the extended objects for conversion to $M$ and classify the objects according to whether $P_{\rm 3D}$ is higher than 50\% or not (``Inside'' or ``Outside'' 3D skymap, respectively).
The objects classified as ``Outside'' are likely to be unrelated to the GW event.
If both \texttt{rMeanPSFMag} and \texttt{iMeanPSFMag} are set to $-999$, the $P_{\rm 3D}$ are not evaluated, and these objects are classified as ``No Info.''

\begin{figure*}
	\begin{center}
		\includegraphics[width=\hsize]{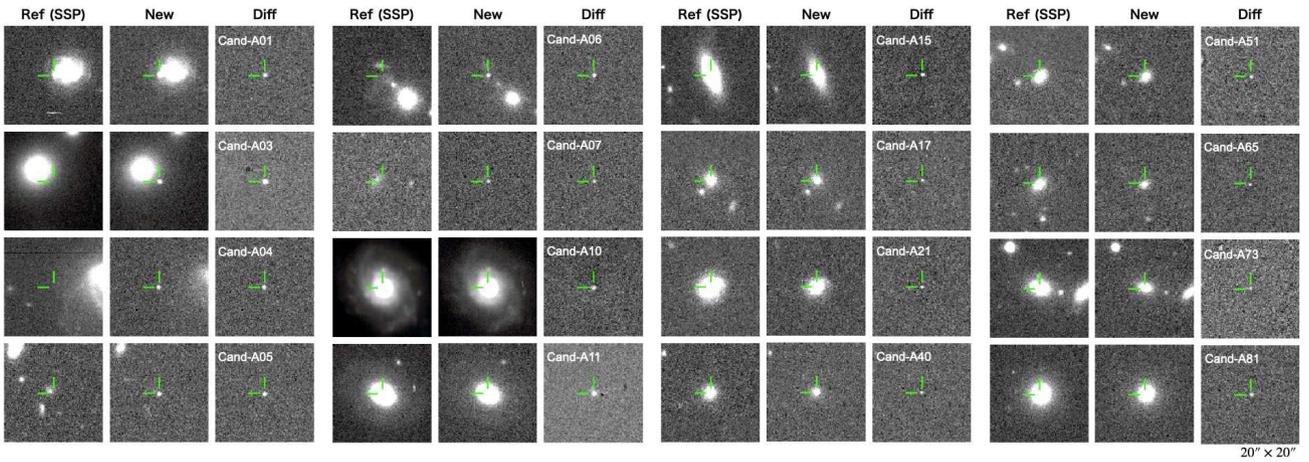} 
	\end{center}
	\caption{Examples of the candidates obtained in the fields with HSC-SSP reference images with the image subtraction; HSC-SSP reference images (Ref), our observation images (New) and difference images (Diff).}
	\label{fig:Candidates_SSP}
\end{figure*}

Finally, we perform a visual inspection because bogus detections remain in these candidates.
After the visual exclusion of bogus detections, we finally obtain 83 candidates.
Figure \ref{fig:Candidates_SSP} shows some example images of these candidates.
A detailed information of these candidates is shown in Table \ref{tab:candidates-SSPref} (``Off-center'' and ``No close objects'') and Table \ref{tab:candidates-GC} (``Center of extended object'').
Since only Cand-A10 has a high $P_{\rm 3D}$, we conclude this source as a final candidate of an electromagnetic counterpart of S190510g. However, this source may result from a variability of an active galactic nucleus because it is located at center of the extended object. We cannot evaluate $P_{\rm 3D}$ of three candidates (Cand-A07, Cand-A08 and Cand-A09) because these have no close extended object and thus cannot rule out their possibility to be the counterpart of S190510g.
 
\begin{table*}
	\tbl{Candidates obtained in the fields with the HSC-SSP reference images with the image subtraction (Off center of the extended objects or No close objects).}{%
	\begin{tabular}{cccccc} \hline
		Name & R.A. & decl. & Mag.\footnotemark[$\dag$] & $\theta_{\rm sep}$\footnotemark[$\ddag$] & $P_{\rm 3D}$ \\
		 & (J2000) & (J2000) & (AB) & [arcsec] & [\%] \\
		\hline \hline
		\multicolumn{6}{c}{Off-center (Outside\footnotemark[$*$])}\\ \hline
		Cand-A01 & $13^h38^m44^s.73$ & $-01^{\circ}51'33''.0$ & 21.21 & 2.9 & 13.4 \\
		Cand-A02 & $13^h31^m44^s.51$ & $-00^{\circ}16'03''.3$ & 22.14 & 14.3 & 2.9 \\
		Cand-A03 & $13^h40^m17^s.42$ & $+00^{\circ}01'54''.8$ & 20.16 & 4.2 & 43.3 \\
		Cand-A04 & $13^h27^m22^s.92$ & $+01^{\circ}28'19''.2$ & 20.57 & 13.0 & 31.8 \\
		Cand-A05 & $13^h34^m33^s.39$ & $+00^{\circ}57'35''.6$ & 20.41 & 11.7 & 6.0 \\
		Cand-A06 & $13^h31^m45^s.48$ & $-00^{\circ}04'26''.7$ & 21.17 & 6.8 & 10.3 \\
		\hline
		\multicolumn{6}{c}{No close objects}\\ \hline
		Cand-A07 & $13^h29^m04^s.60$ & $+00^{\circ}14'19''.7$ & 21.43 & --- & --- \\
		Cand-A08 & $14^h44^m09^s.04$ & $+04^{\circ}43'51''.7$ & 20.48 & --- & --- \\
		Cand-A09 & $14^h54^m15^s.63$ & $+04^{\circ}47'35''.0$ & 17.50 & --- & --- \\
		\hline
	\end{tabular}}\label{tab:candidates-SSPref}
	\begin{tabnote}
		\footnotemark[$\dag$] Magnitudes in the difference image before the Galactic extinction correction. \\
		\footnotemark[$\ddag$] Angular separations from the extended object in the PS1 catalog. \\
		\footnotemark[$*$] Outside 3$\sigma$ region of 3D localization map ($P_{\rm 3D}<50\%$). These are not likely to be related to the GW event S190510g.
	\end{tabnote}
\end{table*}

\begin{longtable}{ccccc||ccccc}
	\caption{Candidates obtained in the fields with the HSC-SSP reference images using the image subtraction (Center of extended object).} \label{tab:candidates-GC}
	\hline
	Name & R.A. & decl. & Mag. & $P_{\rm 3D}$ & Name& R.A. & decl. & Mag. & $P_{\rm 3D}$\\
	  & (J2000) & (J2000) & (AB) & [\%] & & (J2000) & (J2000) & (AB) & [\%]\\
	\hline \hline
	\endfirsthead
	\hline
	Name & R.A. & decl. & Mag. & $P_{\rm 3D}$ & Name& R.A. & decl. & Mag. & $P_{\rm 3D}$\\
	  & (J2000) & (J2000) & (AB) & [\%] & & (J2000) & (J2000) & (AB) & [\%]\\
	\hline \hline
	\endhead
	\hline
	\endfoot
	\endlastfoot
	\multicolumn{10}{c}{Inside ($P_{\rm 3D}\geq50\%$)} \\ \hline
	Cand-A10 & $13^h29^m50^s.37$ & $-01^{\circ}25'44''.5$ & 21.11 & 65.0 &&&&& \\
	\hline
	\multicolumn{10}{c}{Outside ($P_{\rm 3D}\leq50\%$)} \\ \hline
	Cand-A11 & $13^h32^m18^s.73$ & $-01^{\circ}54'19''.1$ & 20.36 & 27.8 & Cand-A48 & $13^h40^m08^s.70$ & $+00^{\circ}29'45''.9$ & 21.85 & 8.6 \\
    Cand-A12 & $13^h31^m37^s.85$ & $-01^{\circ}32'54''.4$ & 21.04 & 10.4 & Cand-A49 & $13^h39^m54^s.54$ & $+00^{\circ}06'52''.3$ & 22.07 & 3.3 \\
    Cand-A13 & $13^h31^m07^s.23$ & $-01^{\circ}31'55''.9$ & 21.53 & 6.3 & Cand-A50 & $13^h39^m55^s.81$ & $+00^{\circ}12'44''.2$ & 21.48 & 5.1 \\
    Cand-A14 & $13^h30^m59^s.43$ & $-01^{\circ}31'54''.9$ & 21.81 & 12.9 & Cand-A51 & $13^h39^m26^s.82$ & $-01^{\circ}27'10''.6$ & 21.98 & 3.7 \\
    Cand-A15 & $13^h29^m51^s.33$ & $-01^{\circ}31'00''.4$ & 21.00 & 22.9 & Cand-A52 & $13^h39^m23^s.65$ & $+00^{\circ}43'20''.1$ & 22.13 & 2.1 \\
    Cand-A16 & $13^h38^m01^s.90$ & $-01^{\circ}44'52''.1$ & 21.57 & 4.8 & Cand-A53 & $13^h39^m38^s.65$ & $+00^{\circ}10'53''.0$ & 22.16 & 2.3 \\
    Cand-A17 & $13^h37^m46^s.70$ & $-01^{\circ}44'12''.3$ & 21.69 & 2.4 & Cand-A54 & $13^h38^m47^s.24$ & $+00^{\circ}39'02''.3$ & 21.60 & 4.6 \\
    Cand-A18 & $13^h37^m26^s.87$ & $-01^{\circ}38'59''.2$ & 21.46 & 3.1 & Cand-A55 & $13^h38^m48^s.40$ & $+00^{\circ}26'22''.9$ & 22.20 & 3.2 \\
    Cand-A19 & $13^h34^m12^s.43$ & $-01^{\circ}56'15''.8$ & 21.22 & 8.5 & Cand-A56 & $13^h38^m26^s.76$ & $+00^{\circ}22'59''.4$ & 21.53 & 12.9 \\
    Cand-A20 & $13^h34^m33^s.25$ & $-01^{\circ}38'25''.3$ & 20.91 & 20.9 & Cand-A57 & $13^h38^m08^s.24$ & $-01^{\circ}15'30''.0$ & 21.17 & 3.8 \\
    Cand-A21 & $13^h34^m32^s.33$ & $-01^{\circ}31'45''.5$ & 21.60 & 7.8 & Cand-A58 & $13^h38^m01^s.29$ & $+00^{\circ}14'09''.3$ & 21.73 & 10.8 \\
    Cand-A22 & $13^h32^m35^s.23$ & $-01^{\circ}35'32''.9$ & 21.28 & 13.9 & Cand-A59 & $13^h36^m57^s.42$ & $-01^{\circ}25'49''.1$ & 21.50 & 5.5 \\
    Cand-A23 & $13^h32^m56^s.13$ & $-01^{\circ}33'17''.1$ & 21.82 & 4.2 & Cand-A60 & $13^h36^m47^s.35$ & $+00^{\circ}21'45''.7$ & 21.67 & 3.3 \\
    Cand-A24 & $13^h40^m12^s.80$ & $-01^{\circ}28'41''.7$ & 21.64 & 2.5 & Cand-A61 & $13^h36^m49^s.12$ & $+00^{\circ}20'57''.3$ & 20.29 & 18.6 \\
    Cand-A25 & $13^h39^m24^s.14$ & $-01^{\circ}41'49''.7$ & 21.57 & 14.2 & Cand-A62 & $13^h36^m11^s.66$ & $+00^{\circ}40'31''.2$ & 21.78 & 4.8 \\
    Cand-A26 & $13^h39^m40^s.05$ & $-01^{\circ}37'28''.9$ & 21.57 & 8.7 & Cand-A63 & $13^h36^m21^s.98$ & $+00^{\circ}12'27''.7$ & 22.07 & 7.7 \\
    Cand-A27 & $13^h39^m26^s.82$ & $-01^{\circ}27'10''.5$ & 22.03 & 3.7 & Cand-A64 & $13^h35^m20^s.53$ & $-01^{\circ}26'23''.6$ & 21.25 & 28.4 \\
    Cand-A28 & $13^h39^m00^s.44$ & $-01^{\circ}44'34''.4$ & 21.53 & 7.5 & Cand-A65 & $13^h35^m55^s.12$ & $+00^{\circ}39'39''.0$ & 22.11 & 1.0 \\
    Cand-A29 & $13^h37^m59^s.18$ & $-01^{\circ}33'11''.5$ & 21.44 & 3.1 & Cand-A66 & $13^h35^m20^s.63$ & $+00^{\circ}07'16''.2$ & 21.75 & 13.6 \\
    Cand-A30 & $13^h27^m36^s.39$ & $+00^{\circ}09'38''.2$ & 21.65 & 8.7 & Cand-A67 & $13^h35^m11^s.63$ & $+00^{\circ}03'26''.2$ & 21.56 & 10.8 \\
    Cand-A31 & $13^h26^m26^s.27$ & $-01^{\circ}08'50''.1$ & 21.61 & 30.0 & Cand-A68 & $13^h40^m47^s.59$ & $-01^{\circ}15'15''.7$ & 22.11 & 4.3 \\
    Cand-A32 & $13^h26^m23^s.86$ & $+00^{\circ}14'27''.1$ & 21.13 & 20.6 & Cand-A69 & $13^h40^m26^s.53$ & $+00^{\circ}00'26''.1$ & 21.61 & 3.1 \\
    Cand-A33 & $13^h26^m26^s.19$ & $+00^{\circ}05'09''.5$ & 22.04 & 5.4 & Cand-A70 & $13^h28^m32^s.61$ & $+01^{\circ}23'25''.5$ & 21.93 & 3.8 \\
    Cand-A34 & $13^h26^m03^s.24$ & $+00^{\circ}27'40''.2$ & 21.31 & 25.2 & Cand-A71 & $13^h26^m47^s.14$ & $+00^{\circ}03'19''.8$ & 21.71 & 10.2 \\
    Cand-A35 & $13^h25^m49^s.45$ & $+00^{\circ}17'23''.9$ & 20.69 & 19.9 & Cand-A72 & $13^h31^m26^s.35$ & $+01^{\circ}07'52''.0$ & 21.86 & 14.2 \\
    Cand-A36 & $13^h25^m48^s.55$ & $+00^{\circ}10'19''.7$ & 21.71 & 10.2 & Cand-A73 & $13^h30^m17^s.02$ & $+01^{\circ}24'38''.3$ & 22.29 & 6.8 \\
    Cand-A37 & $13^h25^m31^s.41$ & $+00^{\circ}21'44''.1$ & 21.44 & 25.5 & Cand-A74 & $13^h40^m26^s.47$ & $+00^{\circ}14'11''.0$ & 21.57 & 4.0 \\
    Cand-A38 & $13^h35^m00^s.69$ & $-01^{\circ}23'12''.5$ & 21.82 & 4.9 & Cand-A75 & $13^h39^m49^s.56$ & $+00^{\circ}21'58''.2$ & 21.70 & 2.0 \\
    Cand-A39 & $13^h34^m13^s.38$ & $+00^{\circ}53'40''.5$ & 22.20 & 9.9 & Cand-A76 & $13^h39^m38^s.71$ & $+00^{\circ}01'03''.3$ & 21.94 & 2.3 \\
    Cand-A40 & $13^h32^m33^s.20$ & $+00^{\circ}45'29''.8$ & 21.78 & 3.3 & Cand-A77 & $13^h37^m31^s.39$ & $+00^{\circ}35'28''.0$ & 20.73 & 27.1 \\
    Cand-A41 & $13^h32^m51^s.60$ & $+00^{\circ}27'51''.8$ & 21.54 & 7.0 & Cand-A78 & $13^h37^m29^s.93$ & $+01^{\circ}06'33''.8$ & 21.54 & 26.2 \\
    Cand-A42 & $13^h40^m26^s.26$ & $+00^{\circ}50'17''.5$ & 21.42 & 5.5 & Cand-A79 & $13^h36^m38^s.98$ & $+01^{\circ}34'01''.4$ & 21.90 & 37.1 \\
    Cand-A43 & $13^h40^m33^s.36$ & $+00^{\circ}42'58''.7$ & 21.91 & 21.1 & Cand-A80 & $14^h32^m58^s.20$ & $+01^{\circ}33'46''.9$ & 21.44 & 6.4 \\
    Cand-A44 & $13^h40^m14^s.80$ & $-01^{\circ}20'37''.0$ & 22.19 & 1.8 & Cand-A81 & $13^h28^m47^s.90$ & $+02^{\circ}01'12''.8$ & 21.31 & 28.2 \\
    Cand-A45 & $13^h40^m12^s.51$ & $-01^{\circ}06'44''.7$ & 21.86 & 2.0 & Cand-A82 & $13^h33^m24^s.72$ & $+01^{\circ}28'04''.9$ & 21.69 & 14.0 \\
    Cand-A46 & $13^h40^m14^s.87$ & $+00^{\circ}58'01''.9$ & 21.91 & 3.0 & Cand-A83 & $14^h32^m44^s.71$ & $+01^{\circ}31'30''.2$ & 21.34 & 3.1 \\
    Cand-A47 & $13^h40^m11^s.90$ & $+00^{\circ}46'43''.2$ & 21.69 & 3.9 &&&&&  \\
	\hline
\end{longtable}

\subsection{Selection in fields without HSC-SSP reference images}
Next, we examine the fields without the HSC-SSP reference images.
We construct the candidate catalog from the single exposure images, rather than using the method in Section 3.1.
Here, we also focus on point source-like transients in the single-exposure images before stacking.
We perform a forced photometry to the single-exposure images with the {\it hscPipe} and select objects by the following criteria:
(i) \texttt{extendedness}\footnote{The \texttt{extendedness} has the Double type; however, it has only two values 1.0 and 0.0 in {\it hscPipe} v4.0.5.} equals to $0.0$ (i.e. point-like object) and (ii) detected twice with an interval of at least 1 hour.
Additionally, we add a criterion: 
(iii) magnitude $<$ 21.3 mag for matching the objects with the PS1 catalog, which corresponds to the 5$\sigma$ depth\footnote{Cited from PanSTARRS1 Quick Facts in https://panstarrs.stsci.edu/} in $y$-band of the PS1 3$\pi$ steradian survey.
Therefore, the search depth is limited to 21.3 mag by the depth of the PS1 catalog.
For finding sources on a bright region of an extended object, the effective search depth can be shallower than isolated sources: this is because the source on extended objects is detected only when the source is bright enough to make a significant local minimum of brightness between the source and the peak of extended source (\cite{Magnier_2020}).
We find 664477 objects satisfying these criteria.

\begin{figure}
	\begin{center}
		\includegraphics[width=7.cm]{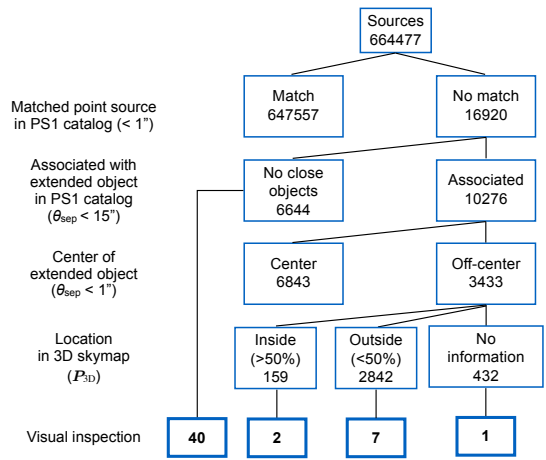} 
	\end{center}
	\caption{Flowchart of the candidate screening and classification process for the selection from the single-exposure images in the fields without HSC-SSP reference images.}
	\label{fig:flowchart_PS1}
\end{figure}

Since these objects include stars, we perform an object screening with criteria similar to the one in Section 3.1.
We show a flowchart for this screening process in Fig. \ref{fig:flowchart_PS1}.
First, we discard 647557 objects positionally coinciding with point-like objects in the PS1 catalog within 1 arcsec, and obtain the remaining 16920 objects.
Next, we classify them based on whether they have PS1 objects which is extended within 15 arcsecs or not.
We obtain 6644 objects without close objects and 10276 objects associated with the extended PS1 objects.
If the objects located at the center of the extended object include non-transient objects, it is difficult to visually classify them without image subtraction.
We discard 6843 objects associated with the extended PS1 objects within 1 arcsec, and then obtain 3433 ``Off-center'' objects.
Considering $P_{\rm 3D}$ of the extended objects associated with the ``Off-center'' objects, 159 objects are classified as ``Inside'' and 2842 objects are classified as ``Outside.''
Both \texttt{rMeanPSFMag} and \texttt{iMeanPSFMag} of the PS1 objects associated with the remaining 432 objects are set to $-999$ in the PS1 catalog.
Therefore, those $P_{\rm 3D}$ are not evaluated, and are classified as `No Information'.

\begin{figure}
	\begin{center}
		\includegraphics[width=8cm]{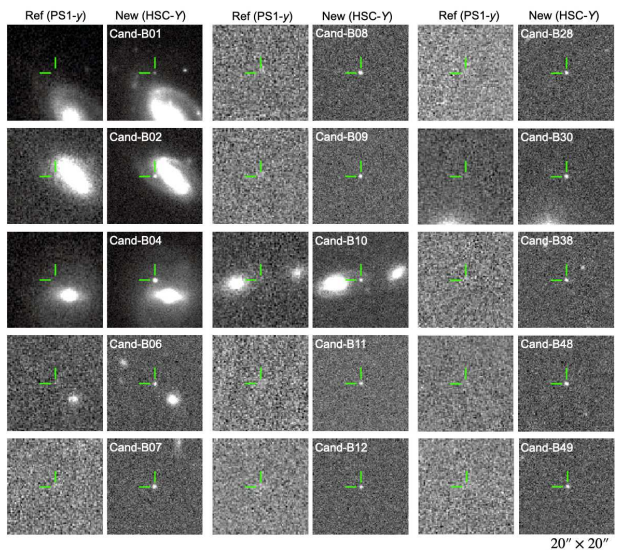} 
	\end{center}
	\caption{Examples of the candidates in the fields without HSC-SSP reference images; reference images in PS1-$y$ band (Ref) and our observation images with HSC (New).}
	\label{fig:Candidates_PS1}
\end{figure}

Finally, we conduct a visual inspection to remove the remaining bogus objects from 10077 objects (6644 ``No close objects,'' 432 ``No information,'' 159 ``Inside,'' and 2842 ``Outside'').
Here, we compare our images with the PS1 stacked images in $g$-, $r$-, $i$-, $z$-, and $y$-band.
We discard the objects when a counterpart below the detection threshold can be recognized in the PS1 catalogs.
These objects in the PS1 images are significantly faint and slightly visible.
We then obtain 50 candidates after the visual inspection as summarized in Table \ref{tab:candidates-Hosted} (``Associated'') and Table \ref{tab:candidates-Hostless} (``No close objects'').
Figure \ref{fig:Candidates_PS1} shows some examples of the candidate.
Since two sources (Cand-B01 and Cand-B02) have high $P_{\rm 3D}$, we also include these sources to final candidates of an electromagnetic counterpart of S190510g. We cannot exclude the possibility to be the counterpart of S190510g for 40 objects tagged ``No close object'' and a ``No information'' because we cannot evaluate $P_{\rm 3D}$  of them.

\begin{table*}
	\tbl{Candidates obtained in the fields without HSC-SSP reference images (Off-center of extended object).}{%
	\begin{tabular}{cccccc} \hline
		Name & R.A. & decl. & Mag. & $\theta_{\rm sep}$ & $P_{\rm 3D}$\footnotemark[$\ddag$] \\
		 & (J2000) & (J2000) & (AB) & [$\,''$] & [\%] \\
		\hline \hline
		\multicolumn{6}{c}{Inside  ($P_{\rm 3D}\geq50\%$)}\\ \hline
		Cand-B01 & $13^h46^m52^s.14$ & $+03^{\circ}45'01''.4$ & 21.24 & 9.96 & 84.8 \\
		Cand-B02 & $13^h44^m36^s.38$ & $+03^{\circ}17'19''.6$ & 20.56 & 3.61 & 73.3 \\
		\hline
		\multicolumn{6}{c}{Outside  ($P_{\rm 3D}\leq50\%$)}\\ \hline
		Cand-B03 & $13^h24^m09^s.46$ & $-05^{\circ}46'17''.6$ & 21.19 & 14.4 & 1.6 \\
		Cand-B04 & $13^h45^m28^s.41$ & $-06^{\circ}01'47''.8$ & 19.47 & 4.41 & 46.7 \\
		Cand-B05 & $13^h32^m58^s.25$ & $+04^{\circ}32'17''.0$ & 21.24 & 13.1 & 2.2 \\
		Cand-B06 & $14^h29^m23^s.39$ & $+04^{\circ}11'39''.1$ & 21.29 & 4.96 & 8.2 \\
		Cand-B07 & $13^h46^m40^s.32$ & $-04^{\circ}54'32''.3$ & 20.99 & 12.5 & 2.1 \\
		Cand-B08 & $13^h34^m46^s.22$ & $+03^{\circ}32'27''.8$ & 21.09 & 12.6 & 2.1 \\
		Cand-B09 & $13^h32^m58^s.25$ & $+04^{\circ}32'18''.0$ & 21.25 & 13.1 & 2.2 \\
		\hline
		\multicolumn{6}{c}{No Information}\\ \hline
		Cand-B10 & $13^h35^m07^s.75$ & $+03^{\circ}39'10''.8$ & 21.00 & 6.66 & --- \\
		\hline
	\end{tabular}}\label{tab:candidates-Hosted}
	\begin{tabnote}
		\footnotemark[$\ddag$] Candidates classified as ``No Information'' are not evaluated $P_{\rm 3D}$, because both \texttt{rMeanPSFMag} and \texttt{iMeanPSFMag} are set to $-999$.
	\end{tabnote}
\end{table*}

\begin{table*}
	\tbl{Candidates obtained in the fields without HSC-SSP reference images (No close objects).}{%
	\begin{tabular}{cccc||cccc}
	\hline
	 Name & R.A. & decl. & Mag. & Name & R.A. & decl. & Mag.\\
	  & (J2000) & (J2000) & (AB) &  & (J2000) & (J2000) & (AB)\\
	\hline \hline
	Cand-B11 & $13^h51^m37^s.87$ & $-01^{\circ}33'35''.8$ & 21.12 & Cand-B31 & $13^h36^m48^s.15$ & $-06^{\circ}00'21''.3$ & 21.00\\
	Cand-B12 & $13^h50^m35^s.39$ & $-01^{\circ}15'03''.6$ & 21.19 & Cand-B32 & $13^h35^m54^s.61$ & $-05^{\circ}08'34''.7$ & 21.18\\
	Cand-B13 & $13^h48^m59^s.87$ & $+00^{\circ}51'10''.1$ & 21.28 & Cand-B33 & $13^h34^m47^s.86$ & $-05^{\circ}17'15''.7$ & 21.23\\
	Cand-B14 & $13^h48^m07^s.72$ & $-01^{\circ}17'11''.2$ & 21.23 & Cand-B34 & $13^h34^m47^s.98$ & $-04^{\circ}53'23''.9$ & 21.20\\
	Cand-B15 & $13^h47^m17^s.40$ & $-01^{\circ}08'05''.5$ & 21.08 & Cand-B35 & $14^h49^m38^s.82$ & $+05^{\circ}50'59''.3$ & 21.22\\
	Cand-B16 & $13^h38^m31^s.49$ & $-04^{\circ}05'43''.1$ & 21.11 & Cand-B36 & $14^h44^m26^s.11$ & $+05^{\circ}08'30''.3$ & 21.24\\
	Cand-B17 & $13^h33^m31^s.76$ & $-04^{\circ}19'40''.6$ & 21.12 & Cand-B37 & $13^h47^m13^s.27$ & $+02^{\circ}29'33''.3$ & 20.83\\
	Cand-B18 & $13^h33^m38^s.84$ & $-04^{\circ}02'48''.8$ & 21.22 & Cand-B38 & $13^h46^m21^s.73$ & $+02^{\circ}18'10''.3$ & 21.28\\
	Cand-B19 & $13^h46^m27^s.30$ & $+05^{\circ}05'20''.9$ & 20.96 & Cand-B39 & $13^h46^m29^s.26$ & $+02^{\circ}54'29''.2$ & 21.08\\
	Cand-B20 & $13^h46^m19^s.13$ & $-03^{\circ}39'49''.7$ & 21.21 & Cand-B40 & $13^h43^m09^s.99$ & $+04^{\circ}09'24''.3$ & 21.17\\
	Cand-B21 & $13^h44^m32^s.57$ & $-03^{\circ}31'19''.6$ & 20.94 & Cand-B41 & $13^h38^m59^s.90$ & $+03^{\circ}32'41''.7$ & 21.15\\
	Cand-B22 & $13^h44^m56^s.75$ & $-03^{\circ}28'32''.5$ & 21.10 & Cand-B42 & $13^h44^m59^s.34$ & $+00^{\circ}01'17''.9$ & 21.18\\
	Cand-B23 & $13^h30^m51^s.69$ & $+02^{\circ}54'48''.7$ & 21.17 & Cand-B43 & $13^h41^m28^s.90$ & $+02^{\circ}30'54''.4$ & 21.25\\
	Cand-B24 & $13^h30^m25^s.64$ & $-05^{\circ}21'44''.8$ & 21.24 & Cand-B44 & $13^h32^m45^s.77$ & $-02^{\circ}03'53''.2$ & 20.98\\
	Cand-B25 & $13^h42^m21^s.53$ & $+04^{\circ}58'35''.3$ & 21.17 & Cand-B45 & $13^h37^m56^s.49$ & $+03^{\circ}04'31''.1$ & 21.21\\
	Cand-B26 & $13^h28^m58^s.19$ & $-03^{\circ}41'20''.8$ & 21.02 & Cand-B46 & $13^h37^m37^s.23$ & $+03^{\circ}29'21''.5$ & 21.10\\
	Cand-B27 & $13^h26^m38^s.58$ & $-04^{\circ}17'45''.9$ & 21.27 & Cand-B47 & $13^h37^m11^s.90$ & $+03^{\circ}32'12''.5$ & 21.18\\
	Cand-B28 & $14^h25^m14^s.70$ & $+04^{\circ}18'08''.1$ & 21.11 & Cand-B48 & $13^h37^m36^s.64$ & $+03^{\circ}33'24''.9$ & 21.17\\
	Cand-B29 & $13^h42^m48^s.55$ & $+00^{\circ}53'02''.9$ & 20.56 & Cand-B49 & $13^h33^m45^s.02$ & $+03^{\circ}21'41''.0$ & 21.13\\
	Cand-B30 & $13^h38^m31^s.49$ & $-04^{\circ}05'43''.1$ & 21.11 & Cand-B50 & $13^h33^m51^s.39$ & $+03^{\circ}22'12''.2$ & 21.02\\
	\hline
\end{tabular}}\label{tab:candidates-Hostless}
\end{table*}

\section{Discussion}
\subsection{Contamination from supernovae}
The candidates include objects unrelated to the GW event, such as supernova (SN).
However, it is difficult to determine their nature because we have only 1-day photometric observations.
Thus, we compare our results with the expected number of SN detections and consider the contamination from them.
As shown below, this comparison demonstrates that our sample is dominated by SNe.
Here, we adopt a similar method to that introduced in \citet{2014PASJ...66L...9N}.
The expected number is estimated by summing mock-SN samples brighter than magnitudes corresponding to the limiting magnitude at redshift weighted with cosmological histories of SN rates.
We assume Type-Ia SN rate of \citet{2014PASJ...66...49O} and core-collapse SN (Ib, Ic, IIL, IIP and IIn) rates of \citet{2012ApJ...757...70D}.
The SN light curves are generated from SN-spectrum evolutions provided by \citet{2007ApJ...663.1187H} for Type-Ia SN.
For core-collapse SN, we generate the light curves from the templates provided by \citet{2002PASP..114..803N}\footnote{The templates including the core-collapse SNe are publicly available at the website of P. Nugent (https://c3.lbl.gov/nugent/nugent\_templates.html).}.
The luminosity distributions of SNe are taken from \citet{2012ApJ...745...31B} and \citet{2012ApJ...757...70D}.
We sample SNe whose brightness is rising in $y$-band assuming the reference images are taken 500 days before the detection.
Although the $Y$-band bandpasses are different between HSC and PS1, we assume same filters for two observations in this estimation.
The effect of the difference is negligible compared to other effects of model uncertainties.

\begin{figure}
	\begin{center}
		\includegraphics[width=7.cm]{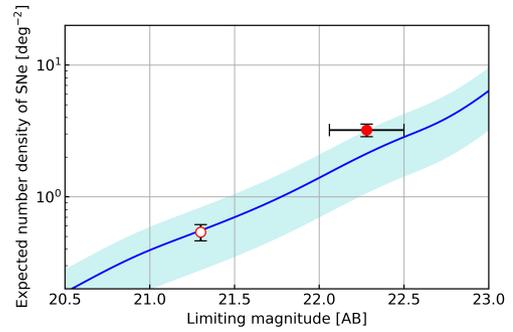} 
	\end{center}
	\caption{Expected number density of SN detections in $Y$-band. The vertical axis indicates the area number density of SNe brighter than the limiting magnitudes in the horizontal axis. The cyan region is the $\pm 50\%$ error originating from the uncertainty of the core-collapse-SN rate density. The dots with error bars are obtained by the number of candidates obtained in the fields with/without the HSC-SSP reference images ({\it filled/open circle}).}
	\label{fig:SNrate}
\end{figure}

Under the above conditions, we derive the expected number density as a function of the limiting magnitude (Fig. \ref{fig:SNrate}).
The cyan region represents the $\pm 50\%$ error originating from the uncertainty of the core-collapse SN rate density.
The dots with error bars are given by $N/S_{\rm field}$, where $S_{\rm field}$ is the area corresponding to the fields with or without the HSC-SSP reference images ({\it filled} or {\it open circle}), and $N$ is the number of candidates found in each field.
The horizontal error bar of {\it filled circle} represents the $1\sigma$-standard-deviation range of the limiting magnitudes in the difference images.
The vertical error bars represent the uncertainties defined as $\sqrt{N}/S_{\rm field}$ by assuming that the number of candidates follows a Poisson distribution with the expected value of $N$.
Although there is a probability that we overlooked some SNe because we neglected the objects located at the center of extended objects in the fields without the HSC-SSP reference images, these dots are within the error region and consistent with the simulation.
Therefore, most of our candidates are likely to be SNe.
The Dark Energy Survey also performed a follow-up observation covering 84 deg$^2$ (65\% of the total probability region) using a DECam with the depths of 20.58 mag ($z$-band), 21.72 mag ($r$-band), and 21.67 mag ($g$-band) and reported that the result of their all candidates are consistent with supernovae (\cite{2020arXiv200700050D}).

\subsection{Comparison with kilonova model and future prospects}
In this section, we compare our search depths with a kilonova model, and discuss future prospects for the follow-up observations with Subaru/HSC.
When we convert the absolute magnitudes to the apparent magnitudes, we correct the Galactic extinction by assuming $E(B-V)=0.10$ mag (\cite{2011ApJ...737..103S}).

\begin{figure}
	\begin{center}
		\includegraphics[width=7.8cm]{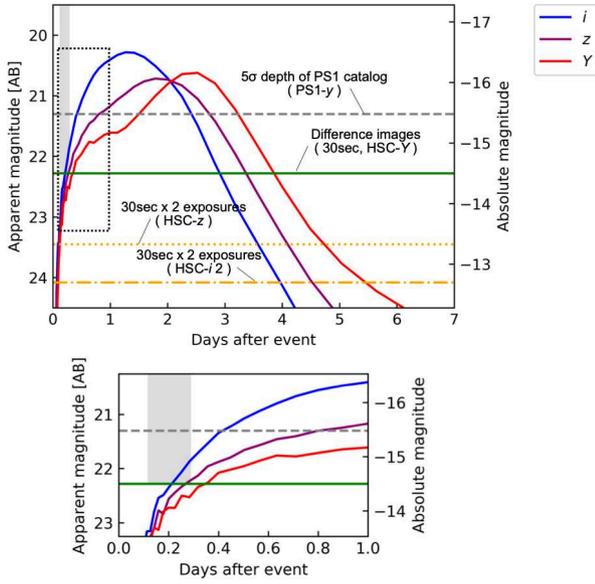} 
	\end{center}
	\caption{Multi-color ($i$-, $z$- and $Y$-band) light curve of the kilonova model with an ejecta mass $M_{\rm ej}=0.05M_{\odot}$ and an electron fraction $Y_{\rm e}=0.30-0.40$ (The $i$- and $z$-band cases are shown in \citet{2020ApJ...901...29B}), assuming that it is located at the distance of 227 Mpc reported in S190510g. Horizontal lines indicate the limiting magnitude in the difference images ({\it green solid line}), the 5$\sigma$ depth of PS1 $y$-band catalog ({\it gray dots line}) and the limiting magnitudes in HSC-$z$ or HSC-$i2$ band calculated by HSC Exposure Time Calculator under the assumption of 30 s $\times$ 2 exposures ({\it orange dotted lines}), respectively. A gray-filled square shows a range of time and depth we observed. A bottom panel is the enlarged figure of the dashed-frame part in the upper panel.}
	\label{fig:Comparison}
\end{figure}

Here, we adopt a kilonova based on the radiative transfer simulations by \citet{2020ApJ...901...29B} for comparison with the early phase of brightness evolution.
We assume that a kilonova with the same propertieis with GW170817 is located at the distance of 227 Mpc reported in S190510g.
Figure \ref{fig:Comparison} shows multi-color ($i$-, $z$- and $Y$-band) light curve of a kilonova model with an ejecta mass $M_{\rm ej}=0.05M_{\odot}$ and an electron fraction $Y_{\rm e}=0.30-0.40$ (no lanthanide).
This parameter set can explain the observed early multi-color light curve of AT2017gfo.
Properties of the light curves are affected by choice of the ejecta mass and electron fraction.
The peak bolometric luminosity roughly scales with the Mej to the power of 0.35 (e.g. \cite{Fern_ndez_2016, Tanaka_2016, Metzger_2019}).
Also, the electron fraction Ye influences the light curves through the opacity: for example, if the $Y_{\rm e}$ is low ($Y_{\rm e}<0.25$), the ejecta becomes Lanthanide rich and the opacity becomes higher.
In such case, the light curves can be fainter by 2-3 mag than our fiducial model ($Y_{\rm e}=0.30-0.40$) in the early time.

For observational limiets, we show the limiting magnitude in the difference images ({\it green solid line}), the 5$\sigma$ depth of the PS1-$y$ band catalog ({\it gray dots line}) and the limiting magnitudes in HSC-$z$ or HSC-$i2$ band ({\it orange dots lines}), respectively, by horizontal lines.
The values in HSC-$z$ and HSC-$i2$ are calculated with HSC ETC by assuming an exposure time of 60 s (30 s $\times$ 2 exposures).
We can observe an area of $\sim$60 deg$^2$ in both $z$- and $i$- bands with this exposure time during a half night of the telescope time.
We note that the limiting magnitudes could be shallower ($\sim$0.5 mag) depending on the sky condition as in the case of our $Y$-band observation.

This comparison demonstrates that observations using Subaru/HSC can detect the kilonova emission in $i$-, $z$-, and $Y$-bands during peak times even at 227 Mpc.
For S190510g, our observations covered approximately 0.1-0.3 days (2.8-7.0 hours), as shown in the gray shaded area in Fig. \ref{fig:Comparison}.
If the emission is purely powered by radioactive decays, the emission still rises in those phases, as shown in Fig. \ref{fig:Comparison}.
However, if ejecta are further heated by cocoon produced by the interaction between the relativistic jets and the ejecta, the emission might be brighter than this kilonova model (\cite{Arcavi_2018}).
Therefore, the early observations will be important to provide constraints on the emission models.

Observations with the Subaru/HSC can detect kilonova fainter than GW170817/AT2017gfo even at the distance of S190510g (227 Mpc).
For example, if the $M_{\rm ej}$ is 0.01$M_{\odot}$ with all the other parameters fixed, the peak bolometric luminosity is decreased from our fiducial model by 40\%.
Then, the peak of  the light curves gets $\sim$0.6 mag fainter under an assumption that the magnitude in each band is decreased following the bolometric luminosity.
Subaru/HSC can easily detect such an emission from the rising phase.
Also, if the Ye is low ($Y_{\rm e}<0.25$), the magnitudes are expected to be fainter by 2-3 mag in the early time as mentioned above.
Even for such a case, we can also detect the peak of the light curve in $i$- and $r$-bands.

Here, we also emphasize the importance of the deep reference image. The horizontal solid line shows our limiting magnitude for the fields with HSC-SSP reference image, while the dashed line shows the limit of search depth in the fields without the reference.
In the latter case, observations are limited by the depth of the PS1 images, and thus, we cannot detect transient sources fainter than 21.3 mag.
For this particular kilonova model at 227 Mpc distance, if the deep reference image is available, the first detection in $Y$-band would be 0.4 days compared to 1.5 days without the deep reference image.

\begin{figure}
	\begin{center}
		\includegraphics[width=8.5cm]{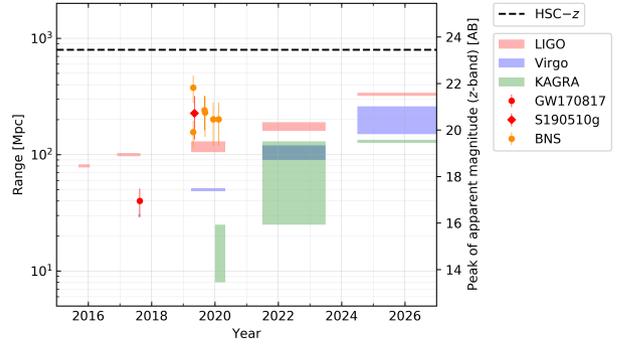} 
	\end{center}
	\caption{BNS range during each observing run shown in the document of observing run plans (https://dcc.ligo.org/public/0161/P1900218/002/SummaryForObservers.pdf). Dots with error bar are the distance (left vertical axis) of GW events reported in O1, O2 and O3. The right vertical axis refers to a peak magnitude of the light curve in $z$-band in the same kilonova model as Fig. \ref{fig:Comparison} on an assumption that it is located at a distance shown in the left vertical axis. Black dashed line indicates the expected limiting magnitude in HSC-z band.}
	\label{fig:BNSrange}
\end{figure}

Finally, we consider future prospects in the fourth and fifth observing runs (O4 and O5) of the GW interferometers.
Colored squares in Fig. \ref{fig:BNSrange} correspond to a BNS range and an observation period in the each observing run shown in the document of observing run plans\footnote{https://dcc.ligo.org/public/0161/P1900218/002/SummaryForObservers.pdf}.
Dots with error bars are distances (left vertical axis) of GW events reported in O1, O2, and O3.
The right vertical axis refers to a peak magnitude of the light curve in $z$-band in the same kilonova model as Fig. \ref{fig:Comparison} on an assumption that it is located at a distance shown in the left vertical axis.
The limiting magnitude evaluated with the exposure time of 60 s in HSC-$z$ band is 23.45 mag ({\it black dashed line}). We expect that observations with HSC will sufficiently attain the BNS ranges expected in O4 or O5.

Here, we estimate how many survey observations for GW events classified as BNS can be performed with Subaru/HSC.
Assuming the Subaru telescope can view the sky above $20^{\circ}$ of elevation, the total area of visible sky is approximately 90\% of the whole sky in a day.
Since the nature seeing becomes poor in the elevations lower than $20^{\circ}$, they are undesirable for the good imaging quality.
Furthermore, typical hours of ``night'', 8 hours, reduces this ratio to approximately 30\%.
In addition, considering nights available for HSC ToO observations, which is estimated as 30\% (\cite{Utsumi:2017snm}), the resulting chance we can conduct the follow-up observation will be approximately 9\% of BNSs, of which the most probable point is located in the observable sky for HSC.
Assuming the BNS rate of 110-3840 Gpc$^{-3}$yr$^{-1}$ shown in \citet{2019PhRvX...9c1040A} and the mean ranges of LIGO during each observing run, the numbers of BNS-merger events expected to be detected by LIGO are $0.8\sim28.2$ (in O3), $4.9\sim172.4$ (in O4) and $41.4\sim1445.1$ (in O5).
Therefore, the expected number of BNS events that HSC can contribute to is $0.1\sim2.5$ (in O3), $0.4\sim15.2$ (in O4) and $3.7\sim127.7$ (in O5).
These numbers are further increased if we include poorly localized events whose most probable point locates outside the observable sky. It is, therefore, important to prioritize to the events with a better localization.

\section{Summary \& conclusion}
The GW detection S190510g, that may include NSs, had been reported by the LIGO/Virgo Collaboration on May 10, 2019.
For this event, we performed a ToO observation with Subaru/HSC in $Y$-band for the optical-counterpart survey as early as 1.7 hours after the issue of its preliminary alert.
Our observation area, which was selected from the preliminary localization skymap, covers 118.8 deg$^{2}$.
It corresponds to 11.6\% of the total probability in the localization skymap released in the preliminary alert, and 1.2\% in the updated skymap.
We searched for an optical-counterpart by dividing the observed area into two fields, depending on whether a previous reference HSC image is available.

For the fields with HSC-SSP reference images, we searched for optical counterpart by using the image subtraction.
We obtained 83 candidates through screening sources in the difference images.
For the fields without HSC-SSP reference images, we searched our individual observation images by matching the observed sources with PS1 catalog, and found 50 candidates except the sources located at the center of extended object.
We, then, estimate their distance with photometry of associated extended objects.
Finally, we concluded three sources (Cand-A10, Cand-B01 and Cand-B02) as final candidates of the electromagnetic counterpart of S190510g because these candidate are likely located inside the 3D skymap.
We could not rule out the possibility that 44 candidates are related to the GW event because their distance cannot be estimated.
Unfortunately, no spectroscopic observations for them are performed.
The search depth for the second method is shallower than that for the first method because we chose only brighter source than 21.3 mag to match the sources with the PS1 catalog.

We estimated the expected number of SN detections by performing mock observations.
We confirmed that the number density of our candidates was consistent with the expected number within the 50\% uncertainty.
Therefore, it may imply that most of 133 candidates could be SNe.

By comparing with a radioactive kilonova model reproducing AT2017gfo, which is based on the radiative trasfer simulations with realistic opacity (\cite{2020ApJ...901...29B}), we found that our observations were sufficiently deep to detect the kilonova emission well before the peak at $\sim$230 Mpc distance.
We showed that our observations reached the sensitivity to detect the kilonova emission from the future BNS events in O4 and O5.
With the current estimate of the event rate, we find that Subaru/HSC can observe $0.1\sim2.5$ (in O3), $0.4\sim15.2$ (in O4) and $3.7\sim127.7$ (in O5) events.

Our observations demonstrate that follow-up observations with Subaru/HSC can cover over 100 deg$^2$ within a few hours after the GW event.
For future GW observing runs, most of the GW events will be discovered at $>$100 Mpc distance; therefore, deep and wide observations, as shown in this paper, will become more important.
The early observations with Subaru/HSC will enable the early detection of the EM counterpart for relatively distant objects. Such an early detection will facilitate spectroscopic and multi-color photometric follow-up observations, which are crucial to firmly identify of the EM counterpart. Also, since radioactive heating and shock heating models predict different behaviors before 0.5 days, early deep observations as presented in this paper put strong constraints on the emission mechanism of early kilonova emission.


\begin{ack}
We are grateful to the staff of Subaru Telescope for their help in the observations of this work.
This work was supported by MEXT KAKENHI (Grants JP24103003 and JP17H06363), JSPS KAKENHI (Grants JP16H02158, JP26800103, JP19H00694, JP20H00158 and JP17K14255) and the U.S. Department of Energy (Grant DE-AC02-76-SF00515).
This paper is based on data collected at Subaru Telescope, which is operated by the National Astronomical Observatory of Japan.
We would like to thank Editage (www.editage.com) for English language editing.
\end{ack}

%
%

\bibliography{myref}

\end{document}